\documentclass[a4paper]{article}

\usepackage{INTERSPEECH2019}
\usepackage{subcaption}

\def\x{{\mathbf x}}
\def\y{{\mathbf y}}

\title{Investigating the Lombard Effect Influence on End-to-End Audio-Visual Speech Recognition}
\name{Pingchuan Ma$^1$, Stavros Petridis$^{1,2}$, Maja Pantic$^{1,2}$}
\address{
  $^1$Imperial College London \\
  $^2$Samsung AI Center, Cambridge}
\email{\{pingchuan.ma16, stavros.petridis04\}@imperial.ac.uk}

\begin{document}

\maketitle
\begin{abstract}
Several audio-visual speech recognition models have been recently proposed which aim to improve the robustness over audio-only models in the presence of noise. However, almost all of them ignore the impact of the Lombard effect, i.e., the change in speaking style in noisy environments which aims to make speech more intelligible and affects both the acoustic characteristics of speech and the lip movements. In this paper, we investigate the impact of the Lombard effect in audio-visual speech recognition. To the best of our knowledge, this is the first work which does so using end-to-end deep architectures and presents results on unseen speakers. Our results show that properly modelling Lombard speech is always beneficial. Even if a relatively small amount of Lombard speech is added to the training set then the performance in a real scenario, where noisy Lombard speech is present, can be significantly improved. We also show that the standard approach followed in the literature, where a model is trained and tested on noisy plain speech, provides a correct estimate of the video-only performance and slightly underestimates the audio-visual performance. In case of audio-only approaches, performance is overestimated for SNRs higher than -3dB and underestimated for lower SNRs.

\end{abstract}

\section{Introduction}

It is well known that speakers adapt their speaking style in noisy backgrounds in order to make their speech more intelligible. This is known as the Lombard effect \cite{lombard1911signe} and it is acoustically characterised by an increase in the sound intensity, fundamental frequency, vowel duration and a shift in the formant frequencies \cite{junqua1993lombard,pittman2001recognition,alghamdi2018corpus,summers1988effects}. Visually, it is characterised by hyper-articulation \cite{garnier2018hyper,vsimko2016hyperarticulation} and more pronounced rigid-head motion \cite{summers1988effects,vatikiotis2007audiovisual}.

Recently several audio-visual speech recognition models have been  presented \cite{petridis2018end,chung2016lipSentences,sterpu2018attention,petridis2018audio} which aim to augment the performance of acoustic speech recognisers. The main application of such systems is in noisy acoustic environments since the main assumption is that the visual signal is not affected by noise
and can therefore enhance the performance of speech recognition systems. However, this assumption is not true due to the Lombard effect which also affects the lip movements. In addition, such models are usually trained with plain\footnote{The terms plain and non-Lombard speech are used interchangeably in this work.} speech which is artificially mixed with additive noise. This approach does not correspond to a realistic scenario where  Lombard (and not plain) speech is mixed with noise. This mismatch can potentially harm the performance of audio-only, video-only and audio-visual speech recognisers. 

Few works have investigated the impact of the Lombard effect on audio-only speech recognition
\cite{marxer2018impact,junqua1993lombard,wakao1996variability}. The main finding is that the performance of a model trained on plain speech mixed with noise is significantly degraded when tested on noisy Lombard speech. This is true even when compensated Lombard speech is used, i.e., the Lombard utterances are normalised to the same energy as the plain speech utterances, although the performance drop is smaller in this case \cite{marxer2018impact}. A similar performance degradation  has also been reported for speaker recognition  \cite{hansen2009analysis}. However, if noisy Lombard speech is used for training then a significant improvement is reported. It is also worth pointing out that the performance of a model trained and tested on noisy Lombard is higher than a model trained and tested on noisy plain speech \cite{marxer2018impact}. 

Even fewer works have investigated the effect of the Lombard reflex on visual and audio-visual speech recognition and the results are not conclusive. Marxer et al. \cite{marxer2018impact} report an improvement on the recognition of visual Lombard speech no matter if the model is trained on plain or Lombard speech. As expected the improvement is higher when visual Lombard speech is used for training. On the other hand, Heracleous et al. \cite{heracleous2013analysis} reported a performance drop when there is a mismatch between training and testing conditions. The same conclusion was also reached when an audio-visual speech recognition system was used. Finally, it has recently been shown that the mismatch between plain and Lombard speech can also affect the performance of audio-visual speech enhancement models \cite{michelsanti2019effects}.

In this work, we investigate the impact of the Lombard effect on end-to-end audio-only, video-only and audio-visual speech recognition. To the best of our knowledge, this is the first work that studies the Lombard effect within the framework of deep end-to-end models which learn to extract features directly from the raw images and audio waveforms. This is in contrast with the majority of previous works which  used hand-crafted features in combination with Gaussian Mixture Models-Hidden Markov Models (GMM-HMMs).  

\looseness-1
In addition, we also consider both multi-speaker and subject-independent scenarios. The former has been extensively studied in previous works \cite{heracleous2013analysis,marxer2018impact} and offers an insight on the impact of the Lombard effect. However, in a real scenario we are mainly interested in  the performance on unseen speakers. Hence, we first conduct multi-speaker experiments in order to test the claims made by prior works. Then we also conduct subject-independent experiments in order to investigate the performance on unseen speakers which has not been explored before. This is of particular interest since it is known that the degree of the Lombard effect is highly speaker dependent \cite{junqua1993lombard,marxer2018impact}. 

Finally, we report results on sentence-level speech recognition. This is in contrast to previous works which mainly focus either on isolated words \cite{heracleous2013analysis} or on specific words within a sentence \cite{marxer2018impact}. We believe that the conclusions reached by this approach can be more useful for a practical speech recognition system where the goal will most likely be to recognise all words in a sentence rather than recognise just isolated words.

We show that properly modelling Lombard speech during training leads to improved performance for audio-only, video-only and audio-visual speech recognition models in all experiments. We also show that in subject-independent experiments, including even a relatively small set of Lombard speech during training can significantly improve the performance of an audio-visual speech recogniser in real conditions, i.e., when testing on noisy Lombard speech. Finally, we show that the standard approach followed in the literature, where noise is mixed with plain speech for training and testing, overestimates the actual performance of audio-only models on noisy Lombard speech for Signal-to-Noise ratios (SNRs) higher than -3dB but underestimates it for lower SNRs.
On the other hand, the visual performance is correctly estimated in all scenarios and the audio-visual performance is slighlty underestimated.

\section{Lombard Grid Database}
For the purpose of this study, we use the Audio-Visual Lombard Gird corpus \cite{alghamdi2018corpus}. The corpus consists of 5400 utterances from 54 speakers (30 females and 24 males), with 100 utterances (50 Lombard and 50 plain) per speaker. Each utterance is composed of a six word sequence from the combination of the following components: \textless command: 4\textgreater \textless colour: 4\textgreater \textless preposition: 4\textgreater \textless letter: 25\textgreater \textless digit: 10\textgreater \textless adverb: 4\textgreater, where the number of choices for each component is indicated in the angle brackets.

During speaking, both frontal and profile faces were simultaneously recorded at 25 frames per second (fps) and audio was recorded at 48kHz and downsampled to 16kHz.  Recordings for each utterance were collected in two conditions, Lombard (L) and Non-Lombard (NL). The NL condition was performed by reading sentences to a condenser microphone placed 30cm in front of the participants, in which the own-voice attenuation was compensated. The L condition follows the same setting, but speech-shaped noise at 80dB sound pressure level  was presented to participants via headphones.

\section{Architecture}
\begin{figure}[t]
    \centering
    \includegraphics[width=.32\textwidth]{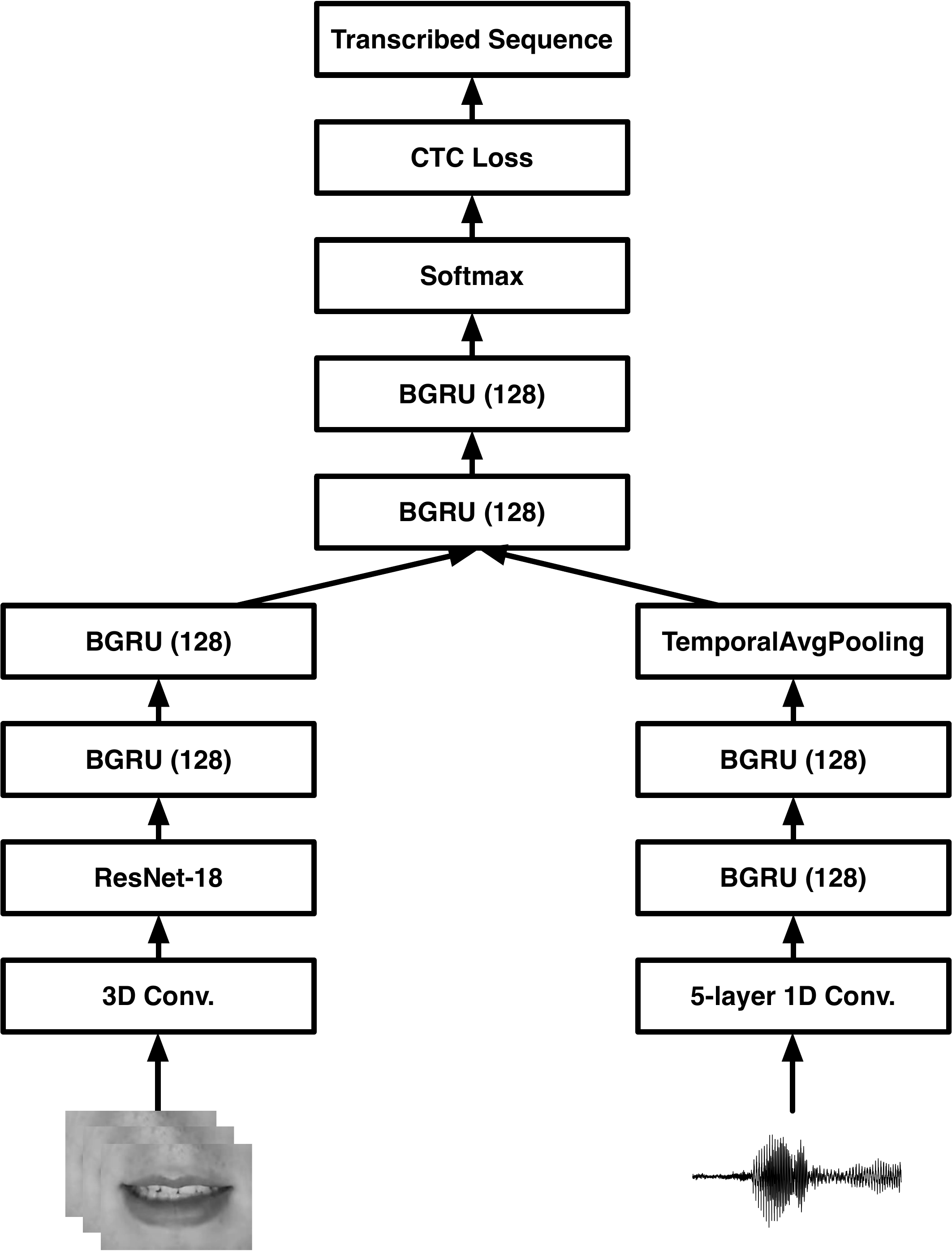} 

    \caption[Audio-visual architecture]{
    End-to-end audio-visual speech recognition architecture overview. Raw images and audio waveforms are fed to the visual and audio streams, respectively, which produce features at the same frame rate at the bottleneck layer. These features are fused together and fed into another 2-layer Bidirectional Gated Recurrent Units (BGRU) to model the temporal dynamics. Connectionist Temporal Classification (CTC) \cite{graves2006connectionist} is used as the loss function.
    }
    \vspace{-.15in}
\label{av_architecture}
\end{figure}
The end-to-end audio-visual speech recognition architecture is shown in Fig. \ref{av_architecture} and is similar to the one proposed in \cite{petridis2018end}. A CTC loss is added so the model can recognise continuous speech. 


\subsection{Visual Stream}
The visual stream  consists of a spatiotemporal convolutional layer, followed by a ResNet-18 \cite{he2016deep} and a 2-layer BGRU. Specifically, the temporal-wise 3D convolutional layer has a kernel size of 5 frames. Then, frame-level features are extracted by ResNet-18. The output of ResNet-18 is fed to a 2-layer BGRU to model the temporal dynamics of visual features. Note that the outputs of the  forward and backward GRU are concatenated together instead of added together. This means that although there are 128 GRU cells, the features produced by the GRU have a dimensionality of 256.

\subsection{Audio Stream}
The audio stream consists of 5 temporal convolutional blocks, followed by a 2-layer BGRU and an average pooling layer. Each convolutional block includes a temporal convolutional layer, ReLU activation and batch normalisation.
The first temporal convolutional layer uses a kernel of 5ms and a stride of 0.25ms to extract fine-scale spectral information. The output of the convolutional layers is fed to a 2-layer BGRU. Similarly to the visual stream, the outputs of the forward and backward BGRUs are concatenated. Finally, an averaging pooling layer is used to reduce the audio frame rate to  the visual frame rate. 

\subsection{Fusion Layers}
Once the 256 audio features and 256 visual features are extracted, they are concatenated and fed into a 2-layer BGRU to model their temporal dynamics. Then a softmax layer follows which provides the characters probabilities for each frame.

\subsection{Connectionist Temporal Classification}
The CTC loss is used to transcribe directly between inputs and target outputs without any intermediate annotation. Given an input sequence $\x=(x_1, ..., x_T)$, CTC sums over the probabilities of all valid alignments with length $T$ to obtain the posterior of the target sequence $\y=(y_1, ..., y_L)$:
$$
p(\y|\x) = \sum_{valid Alignments}\prod_{t=1}^Tp_t(a_t|\x)
$$
where $p_t(a_t|\x)$ is the per time-step probability and the product computes the probability of a single valid alignment. The CTC loss is computed by the negative log likelihood of the posterior probability.

\begin{figure*}
\centering
\begin{subfigure}{.5\textwidth}
  \centering
  \includegraphics[width=.9\textwidth]{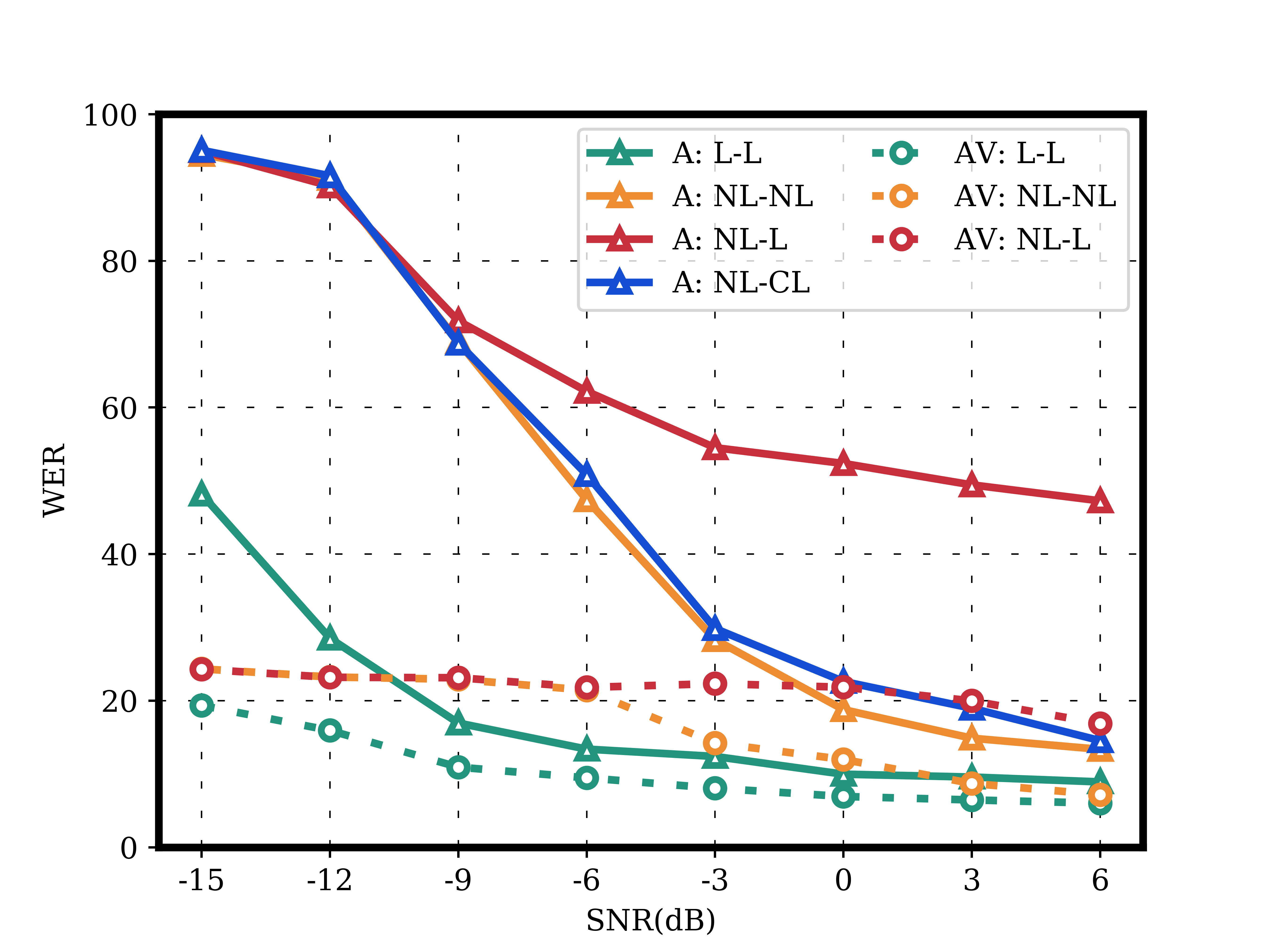}
  \caption[]{
    Case of SNR-specific models.}
  \label{multi_speaker_lombard_effect}
\end{subfigure}%
\begin{subfigure}{.5\textwidth}
  \centering
  \includegraphics[width=.9\textwidth]{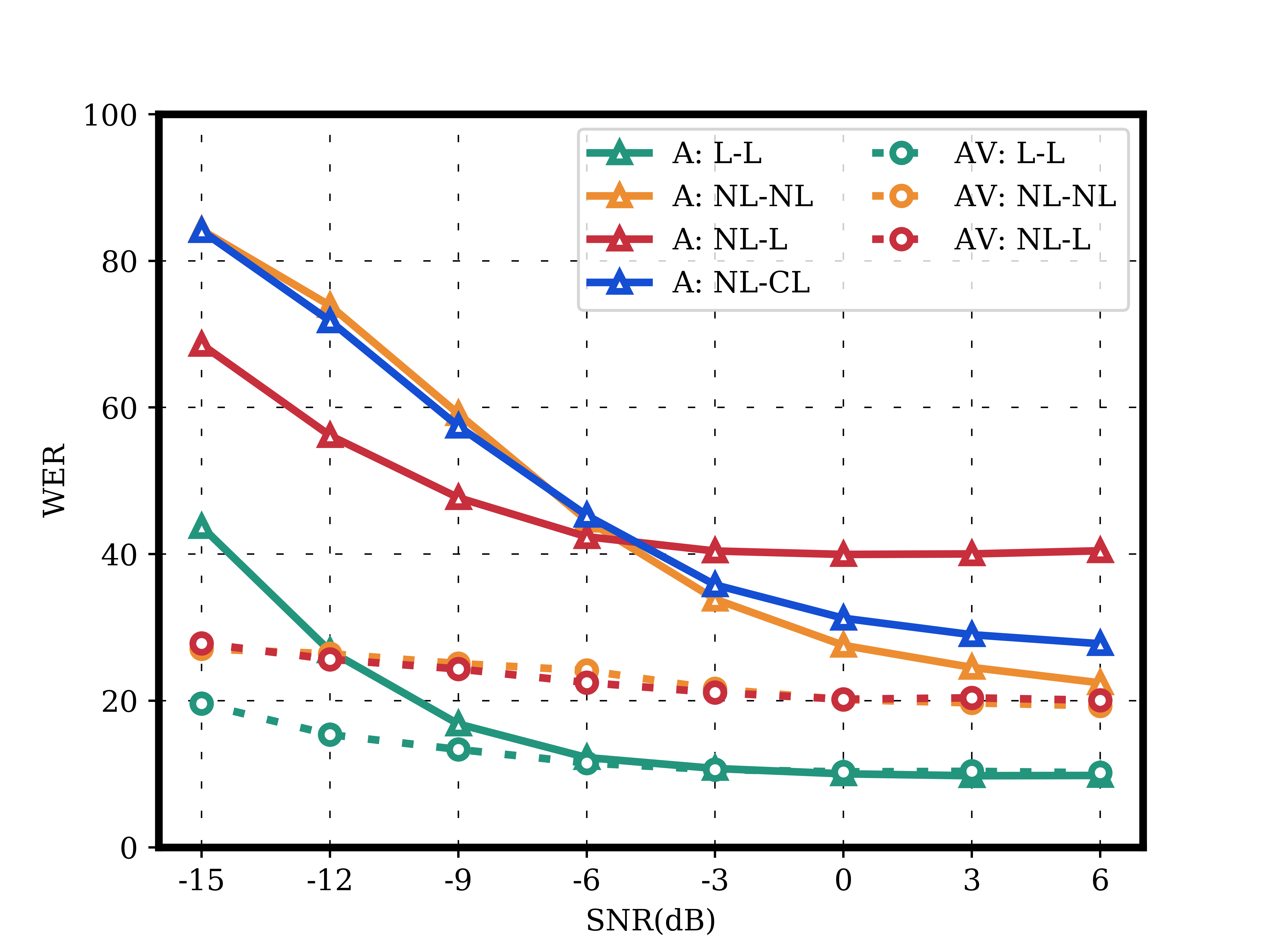}
  \caption{Case of single model trained on all SNRs.}
\label{multi_speaker_SNR_augmented_lombard_effect}
\end{subfigure}
\caption[]{
    WER of the end-to-end models as a function of the noise level in a multi-speaker scenario. A: audio-only model, AV: audio-visual model, L: Lombard, NL: non-Lombard, CL: `compensated' Lombard. X-Y indicates a model trained on X (L or NL) speech and tested on Y (L or NL or CL) speech. Best seen in colour.}
\label{fig:multiSpeaker}
\end{figure*}

\section{Experimental Setup}
\subsection{Preprocessing}
\subsubsection{Video Preprocessing}
We use dlib \cite{king2009dlib} to detect and track facial landmarks for frontal faces and the  face alignment library proposed in \cite{bulat2017far} for profile faces. 
The faces are first aligned using a neutral reference frame in order to normalise them for rotation and size differences. This is performed using an affine transform using 5 stable points, two eyes corners in each eye and the tip of the nose. Then the centre of the mouth is located based on the tracked points and a bounding box of 140 by 200 and 80 by 60 is used to extract the mouth region of interest (ROI) on frontal and profile faces, respectively.

\subsubsection{Audio Preprocessing}
Lombard utterances have greater energy than plain speech utterances so for a given noise level their SNR is higher than noisy plain speech. So similarly to  \cite{marxer2018impact} we also generate `compensated' Lombard speech, where the energy of Lombard speech is normalised to the same energy as plain speech. In this case, the SNR between Lombard and plain utterances is the same for a given noise level.

To remove the artificial variability of the signals caused by the speaker-to-microphone distance, we follow the approach suggested in \cite{marxer2018impact}. We normalise the non-Lombard and `compensated' Lombard signals to the same root mean square (RMS) of 0.05. For the Lombard signals, we set the RMS to $0.05\cdot \bar{x}_{rms}^L/\bar{x}_{rms}^{NL}$, where $\bar{x}_{rms}^L$ and $\bar{x}_{rms}^{NL}$ are the average RMS value on Lombard speech and non-Lombard speech corpus.


\subsection{Data Augmentation}
\label{ssec:DataAugm}

During training, two data augmentation methodologies are performed in raw images, random cropping and horizontal flipping. Specifically, each frontal mouth ROI is randomly cropped to a size of 130 by 190 and each profile mouth ROI is randomly cropped to a size of 75 by 55. During testing, the central patch is cropped. Horizontal flipping with a probability of 0.5 is used to increase the variation on training samples.

Babble noise at different levels is added into the audio waveforms during training. The SNR levels range from -15dB to 6dB with an interval of 3dB. One of the noise levels or the clean signal is selected under a uniform distribution, which enhance robustness to different noise levels. 


\subsubsection{Training}
We first train each stream from scratch. An initial learning rate of 0.001 and a mini-batch of 64 are used for the audio stream and an initial learning rate of 0.0003 and a mini-batch of 10 are used for the visual stream. We train the audio stream for 400 epochs and the visual stream for 120 epochs separately. Once the audio and visual streams have been trained, their weights are fixed and the 2-layer BGRU used for fusion is trained with an initial learning rate of 0.0003 and a mini-batch of 10. Finally, the entire audio-visual model is fine-tuned for another 40 epochs.

\section{Results}


\begin{table}[t]
  \centering
  \begin{tabular}{lccc}
\toprule
Views & L-L & NL-L & NL-NL \\ \midrule
WER (Frontal) & 23.57 &26.05 &25.59\\ \bottomrule
  \end{tabular}
  \caption{Video-only results on a multi-speaker scenario. L: Lombard, NL: non-Lombard. X-Y indicates a model trained on X (L or NL) speech and tested on Y (L or NL) speech.}
  \label{table:multi-speaker}
\vspace{-.22in}

\end{table}
\subsection{Multi-speaker experiments}
\label{ssec:multiSpeakerExp}

In this set of experiments, we investigate the impact of the Lombard effect in a multi-speaker scenario when end-to-end deep models are used for speech recognition. For the purpose of this study, we use 30, 10 and 10 utterances from each subject for training, validation and testing, respectively.
A similar study has been conducted in \cite{marxer2018impact}, but a traditional GMM-HMM approach was followed and SNR-specific models were trained. 

\looseness - 1
 For comparison purposes we first train SNR-specific audio-only models for non-Lombard and Lombard speech similarly to \cite{marxer2018impact}. Results are shown in Fig. \ref{multi_speaker_lombard_effect} and overall are consistent with the results presented in \cite{marxer2018impact}.  We notice that when we train a model on non-Lombard speech and test it on Lombard speech (red solid line), a significant drop in performance compared to testing on non-Lombard speech (orange solid line) is observed between -9dB and 6dB. 
 This is mainly the consequence of the SNR mismatch between Lombard and plain speech.
 However, between -12dB and -15dB, there is no difference between the two training approaches. 
 When we test on `compensated' Lombard speech (blue solid line), the results are still worse than non-Lombard speech (up to 4\%). 
 This indicates that not only the SNR mismatch affects the performance but also the difference in acoustic characteristics between Lombard and non-Lombard speech, to a smaller extent though.
 
 \begin{figure}[tp]
    \centering
    \includegraphics[width=.48\textwidth]{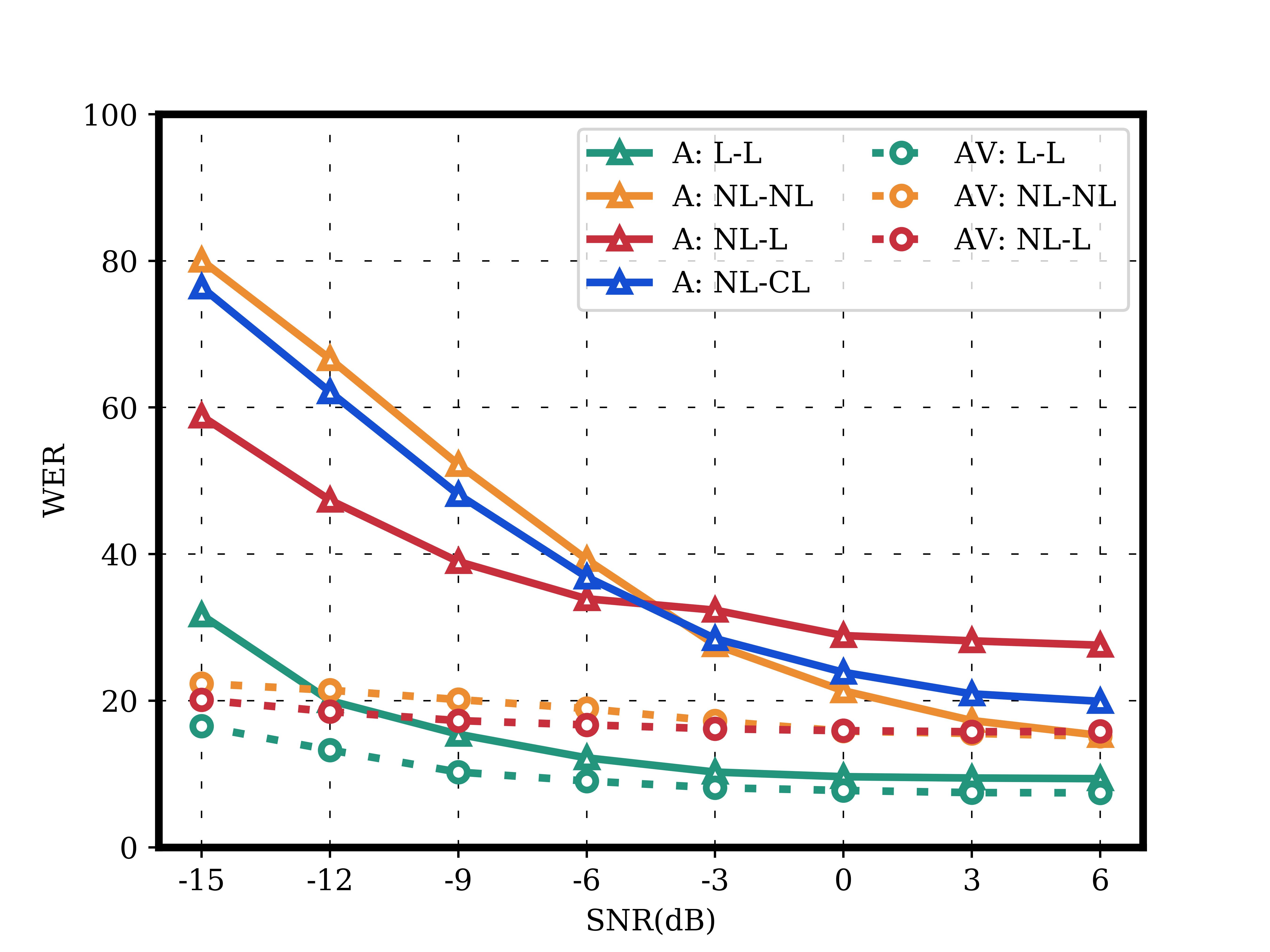}
     \caption[]{WER of the end-to-end as a function of the noise level in a subject-independent scenario. A:  audio-only model, AV: audio-visual model, L: Lombard, NL: non-Lombard, CL: `compensated' Lombard. X-Y indicates a model trained on X (L or NL) speech and tested on Y (L or NL or CL) speech. Best seen in colour.}
\label{subject_independent_lombard_effect}
\end{figure}

\begin{figure}[tp]
    \centering
    \includegraphics[width=.48\textwidth]{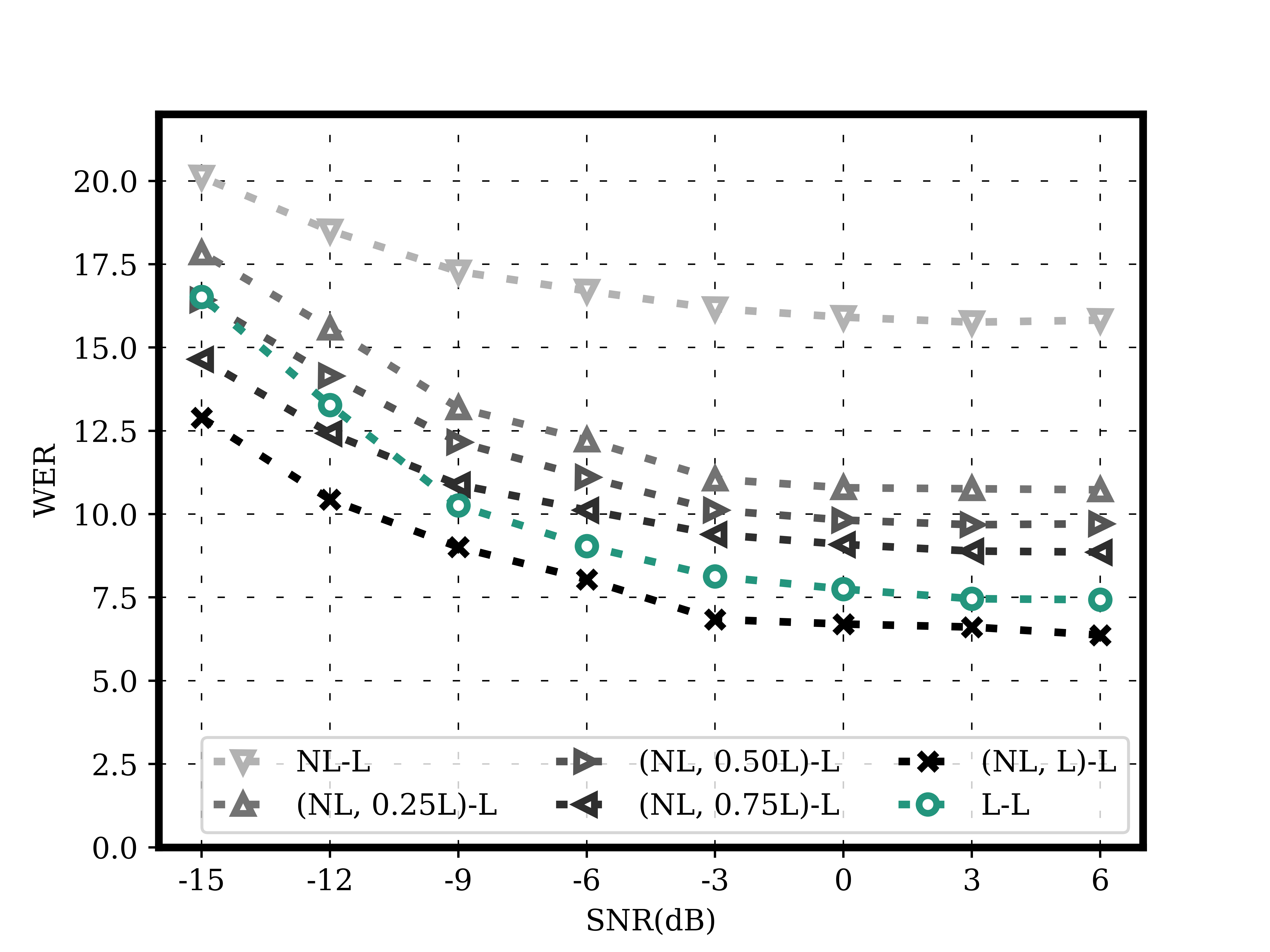}
     \caption[]{WER of the end-to-end audio-visual model as a function of the noise level in a subject-independent scenario. L: Lombard, NL: non-Lombard. (NL,0.25L)-L indicates the performance is reported using a model trained on non-Lombard and 25\% Lombard speech and tested on Lombard speech. The other combinations follow the same pattern. Best seen in colour.}
\label{subject_independent_lombard_effect_perc}
\end{figure}
 

Results for multi-speaker experiments where a single model is trained using the SNR augmentation approach from section \ref{ssec:DataAugm} are shown in Fig. \ref{multi_speaker_SNR_augmented_lombard_effect}. The main difference with the previous set of experiments 
is that the performance on Lombard speech (red solid line), for a model trained on non-Lombard speech, is better than the performance on non-Lombard speech (orange solid line) between -15dB and -6 dB. This is probably due to the fact that during training all SNR levels are seen so the influence of the SNR mismatch between Lombard and plain speech is minimised. The same pattern is also observed for `compensated' Lombard speech (blue solid line). This indicates that although at higher SNRs the performance of a model trained and tested on non-Lombard speech, which is the usual approach in the literature, overestimates the actual performance, in lower SNRs it actually underestimates it. It is also worth pointing out that when we train on Lombard speech, a significant improvement in performance is observed when we test on Lombard speech (green solid line) compared to training on non-Lombard speech and testing either on Lombard (red solid line) or non-Lombard speech (orange solid line).

\looseness - 1
The results of video-only models are reported in Table~ \ref{table:multi-speaker}. A slight improvement of 0.45\% is reported in the case of NL-L over NL-NL. This is not entirely consistent with \cite{marxer2018impact} who reported a greater improvement of 4.6\%.  We also notice that L-L has an absolute improvement of 2.48\% compared to NL-L, which shows the benefit of properly modelling Lombard speech.

The results of audio-visual models are shown in Fig. \ref{multi_speaker_SNR_augmented_lombard_effect}. As expected the audio-visual models have a lower WER compared to audio-only models across all noise levels. It is worth pointing out again that when Lombard speech is properly modelled then a better performance is achieved (green dashed line vs red dashed line).   

\subsection{Subject-independent experiments}

Previous experiments  considered   multi-speaker models. However, in real scenarios, we would like to have a model that works on unseen subjects. To better investigate the impact of Lombard effect in subject-independent experiments, the training, validation and test sets are divided into 36, 6 and 12 subjects, respectively. It is important to note that the same number of female and male speakers are included on validation and test sets.

The results of audio-only experiments are shown in Fig. \ref{subject_independent_lombard_effect}. Similar conclusions to the ones drawn for multi-speaker experiments in Section \ref{ssec:multiSpeakerExp} can be drawn. The performance on Lombard speech (red solid line), for a model trained on non-Lombard speech, is better than the performance on non-Lombard speech (orange solid line) between -15dB and -6 dB.  The same pattern is also observed for `compensated' Lombard speech (blue solid line). Again, this demonstrates that the approach followed in the literature, i.e., training and testing on non-Lombard speech, overestimates the actual performance at higher SNRs but underestimates it in lower SNRs.

\begin{table}[tp]
  \centering
  \begin{tabular}[hbtp]{ lccc}
    \toprule
Views & L-L & NL-L & NL-NL \\ \midrule
WER (Frontal) & 25.00 &27.84 &27.66\\ \midrule
WER (Profile) & 39.45 &47.61 &47.47\\ \bottomrule
  \end{tabular}
  \caption{Video-only results on subject-independent experiments. L: Lombard, NL: non-Lombard. X-Y indicates a model trained on X (L or NL) speech and tested on Y (L or NL) speech.}
  \label{table:subject-independent-experiments}
\vspace{-.22in}
\end{table}

The video-only results are reported in Table \ref{table:subject-independent-experiments}. When we train and test a model on Lombard speech, an absolute improvement of 2.84\% and 8.16\% is observed in frontal faces and profile faces, respectively, over the NL-L scenario. The performance of NL-L is very similar to NL-NL which reveals that the approach followed in the literature (NL-NL) provides a correct estimate of the actual performance (NL-L). We also notice that the performance on profile faces is much worse due to less information being available as well as inaccurate tracking in profile videos.

The results of audio-visual models are shown in Fig. \ref{subject_independent_lombard_effect}. Similarly to the multi-speaker scenario, the best performance is achieved when a model is trained and tested on Lombard speech (green dashed line). It is also clear that training and testing on plain speech (orange dashed line) slightly underestimates the performance of the real scenario, where Lombard speech is used for testing (red dashed line).

Fig. \ref{subject_independent_lombard_effect_perc} shows the performance of an audio-visual model as a function of the percentage of Lombard speech combined with plain speech for training. It is clear that even when the Lombard speech utterances added to the training set account for 25\% of plain speech the gap between NL-L and L-L is reduced to half. Also, when Lombard speech accounts for 50\% of plain speech similar performance to the L-L scenario is achieved for very low SNRs.

\section{Conclusions}
In this work, we investigate the impact of the Lombard effect on audio-only, video-only and audio-visual speech recognition. We show that it is always beneficial to properly model Lombard speech. We also show that training and testing on noisy plain speech, which is commonly used in the literature, is a good estimate for the performance on visual Lombard speech but a bad estimate for the performance of audio-only speech recognition. It would be interesting to investigate in future works how different types of background noise affect the performance of audio-visual speech recognition models.

\section{Acknowledgements}
We gratefully acknowledge the support of NVIDIA Corporation with the donation of the Titan V GPU used for this research. The work of Pingchuan Ma has been funded by the Honda Research Institute.

\bibliographystyle{IEEEtran}

\bibliography{mybib}


\end{document}